\documentclass[12pt]{article}
\usepackage{amssymb,amsmath,epsfig}
\allowdisplaybreaks
\begin{document}
\title{\bf Cosmological Analysis of Scalar Field Models in $f(R,T)$ Gravity}

\author{M. Sharif \thanks{msharif.math@pu.edu.pk} and Iqra Nawazish
\thanks{iqranawazish07@gmail.com}\\
Department of Mathematics, University of the Punjab,\\
Quaid-e-Azam Campus, Lahore-54590, Pakistan.}

\date{}

\maketitle
\begin{abstract}
This paper determines the existence of Noether symmetry in
non-minimally coupled $f(R,T)$ gravity admitting minimal coupling
with scalar field models. We consider a generalized spacetime which
corresponds to different anisotropic and homogeneous universe
models. We formulate symmetry generators along with conserved
quantities through Noether symmetry technique for direct and
indirect curvature-matter coupling. For dust and perfect fluids, we
evaluate exact solutions and construct their cosmological analysis
through some cosmological parameters. We conclude that decelerated
expansion is obtained for quintessence model with dust distribution
while perfect fluid with dominating potential energy over kinetic
energy leads to current cosmic expansion for both phantom as well as
quintessence models.
\end{abstract}

\section{Introduction}

The generic function in $f(R)$ gravity is a coupling-free function
which helps to resolve many cosmological issues. Nojiri and Odintsov
\cite{5} proposed the concept of non-minimal curvature-matter
coupling which introduced a fresh insight among researchers. This
coupling successfully incorporates clusters of galaxies or dark
matter in galaxies, yielding natural preheating conditions
corresponding to inflationary models as well as introduces the idea
of traversable wormholes in the absence of any exotic matter
\cite{4}. Harko et al. \cite{6} proposed a new version of modified
theory whose generic function incorporates curvature as well as
matter known as $f(R,T)$ gravity ($T$ is the trace of
energy-momentum tensor). This function induces strong interactions
of gravity and matter which play a dynamical role to analyze current
cosmic expansion \cite{7}. Sharif and Zubair \cite{10} investigated
some cosmic issues like energy conditions, thermodynamics,
anisotropic exact solutions, reconstruction of some dark energy
models and also studied stability issue in this gravity.

The curiosity for exact solutions of higher order non-linear
differential equations keep motivating the researchers as these are
extensively used to investigate different cosmic aspects. Harko and
Lake \cite{8} discussed exact solutions of cylindrical spacetime in
the presence of non-minimal coupling between $R$ and matter
Lagrangian density ($\mathcal{L}_m$). The higher order non-linear
differential equations of $f(R,T)$ gravity attract many researchers
to construct cosmological analysis via exact solutions of the field
equations. Sharif and Zubair \cite{11} considered exponential and
power-law expansions to evaluate some exact solutions and
kinematical quantities of Bianchi type I (BI) model in this gravity.
Shamir and Raza \cite{9} formulated exact solutions corresponding to
cosmic string as well as non-null electromagnetic field. Shamir
\cite{12} found exact solutions of locally rotationally symmetric BI
model and studied their physical behavior through cosmological
parameters.

In mathematical physics and theoretical cosmology, continuous
symmetry reduces the complexity of non-linear system which
successfully yields exact solutions. In a dynamical system, Noether
symmetry proposes a correspondence between infinitesimal symmetry
generator and conserved quantity. Capozziello et al. \cite{20} used
this approach to find exact solutions of spherically symmetric
spacetime in $f(R)$ gravity. Hussain et al. \cite{23} investigated
the existence of Noether symmetry of power-law $f(R)$ model and
found boundary term to be vanished for flat FRW universe model but
Shamir et al. \cite{24} obtained non-zero boundary term of the same
model. Momeni et al. \cite{a3} explored Noether point symmetry of
isotropic universe in mimetic $f(R)$ and $f(R,T)$ gravity theories.
Shamir and Ahmad \cite{k5} constructed exact solutions in
$f(\mathcal{G},T)$ gravity ($\mathcal{G}$ denotes Gauss-Bonnet
term).

Sanyal \cite{k3} determined exact solutions of Kantowski-Sachs (KS)
universe model through Noether symmetry technique in non-minimally
coupled gravity with scalar field. Camci and Kucukakca \cite{k4}
extended this work by adding BI as well as BIII universe models and
formulated explicit forms of scalar field. Kucukakca et al.
\cite{k1} discussed the presence of Noether symmetry to formulate
exact solutions of locally rotationally symmetric BI universe. Camci
et al. \cite{k2} generalized this work for anisotropic universe
models such as BI, BIII and KS. We have obtained exact solutions of
$f(R)$ power-law model \cite{17} as well as $f(R,T)$ model admitting
indirect non-minimal curvature-matter coupling \cite{16}.

In non-minimally coupled gravitational theory, the Noether symmetry
approach is extensively used to study different cosmological models
and the dynamical role of various scalar field models \cite{18}.
Vakili \cite{22} identified the existence of Noether point symmetry
along with conserved quantity for flat FRW universe and studied the
behavior of effective equation of state (EoS) parameter for
quintessence model in $f(R)$ gravity. Zhang et al. \cite{19}
explored multiple scalar field scenario and formulated a
relationship of potential function with quintessence and phantom
models. Jamil et al. \cite{25} ensured the presence of Noether
symmetry with conservation law for $f(R)$ tachyon model. Sharif and
Shafique \cite{27} obtained exact solutions of isotropic and
anisotropic universe models in scalar-tensor theory non-minimally
coupled with torsion scalar.

In this paper, we discuss the existence of Noether symmetries of
non-minimally coupled $f(R,T)$ gravity interacting with generalized
scalar field model. The format of the paper is as follows. Section
\textbf{2} introduces some basic aspects of this gravity. In section
\textbf{3}, we discuss all possible Noether symmetries with
associated conserved quantities for two particular models of this
theory. We also formulate exact solutions for dust as well as
perfect fluid distribution and study their physical behavior through
some cosmological parameters. In the last section, we present final
remarks.

\section{Some Basics of $f(R,T)$ Gravity}

We consider the action incorporating  gravity, matter and scalar
field as
\begin{equation}\label{1}
\mathcal{I}=\int
d^4x\sqrt{-g}[\mathcal{L}_g+\mathcal{L}_m+\mathcal{L}_{\phi}],
\end{equation}
where $g$ denotes determinant of the metric tensor, $\mathcal{L}_g$
and $\mathcal{L}_{\phi}$ represent gravity and scalar field
Lagrangian densities. For non-minimal coupling, the gravitational
Lagrangian is considered to be a generic function $f(R,T)$ admitting
minimal coupling only with $\mathcal{L}_m$ and $\mathcal{L}_{\phi}$
\cite{6}. In this case, the metric variation of $\mathcal{L}_g$ and
$\mathcal{L}_m$ yields
\begin{eqnarray}\nonumber
&&f_R(R,T)R_{\mu\nu}-\frac{1}{2}f(R,T)g_{\mu\nu}+(g_{\mu\nu}
\nabla_{\mu}\nabla^{\mu}-\nabla_\mu\nabla_\nu)f_R(R,T)+f_T(R,T)
T_{\mu\nu}\\\nonumber&&+f_T(R,T)(g_{\mu\nu}\mathcal{L}_m
-2T_{\mu\nu}-2g^{\mu\nu}\frac{\partial^2\mathcal{L}_m} {\partial
g^{\mu\nu}\partial g^{\mu\nu}})=\kappa^2T_{\mu\nu},
\end{eqnarray}
where subscripts $R$ and $T$ describe corresponding partial
derivatives of $f$, $\nabla_{\mu}$ indicates covariant derivative
and $T_{\mu\nu}$ represents energy-momentum tensor. The divergence
of energy-momentum tensor leads to
\begin{equation}\nonumber
\nabla^{\mu}T_{\mu\nu}=\frac{f_T}{\kappa^2-f_T}\left[(T_{\mu\nu}
+\Theta_{\mu\nu})\nabla^{\mu}\ln
f_T+\nabla^{\mu}\Theta_{\mu\nu}-\frac{g_{\mu\nu}\nabla^{\mu}T}{2}\right].
\end{equation}
In non-minimally coupled modified gravity, the energy-momentum
tensor remains no more conserved. This non-zero divergence
introduces an extra force in equation of motion which is responsible
for deviation of massive test particles from geodesic trajectories.

A generalization of some anisotropic and homogeneous universe models
is given as \cite{fg}
\begin{equation}\label{2}
ds^2=-dt^2+a^2(t)dr^2+b^2(t)(d\theta^2+\zeta(\theta)d\phi^2),
\end{equation}
where $a$ and $b$ are scale factors and $\zeta(\theta)={\theta,~\sin
h\theta,~\sin\theta}$ identify BI, BIII and KS models with the
following relationship
\begin{equation*}
\frac{1}{\zeta}\frac{d^2\zeta}{d\theta^2}=-\xi.
\end{equation*}
For $\xi=0,-1,1$, the spacetime (\ref{2}) corresponds to BI, BIII
and KS universe models, respectively. For perfect fluid, the
energy-momentum tensor is
\begin{equation*}
T_{\mu\nu}=(\rho_m+p_m)u_\mu u_\nu+p_mg_{\mu\nu},
\end{equation*}
where $p_m$ and $\rho_m$ define pressure and energy density,
respectively whereas $u$ represents four-velocity of the fluid. For
the action (\ref{1}), the Lagrangian density of matter and scalar
fields are defined as \cite{ffg}
\begin{equation}\label{3}
\mathcal{L}_m=p_m(a,b),\quad\mathcal{L}_\phi=\frac{\epsilon}
{2}g^{\mu\nu}\partial_{\mu}\phi\partial_{\nu}\phi-V(\phi),
\end{equation}
where $V(\phi)$ denotes potential energy of the scalar field and
$\epsilon=1,-1$ indicate scalar field models, i.e., quintessence and
phantom models.

Phantom model suffers with number of troubles like violation of
dominant energy condition, the entropy of phantom-dominated universe
is negative and consequently, black holes disappear. Such a universe
ends up with a finite time future singularity dubbed as big-rip
singularity \cite{p1}. Different ideas are proposed to cure the
troubles of this singularity such as considering phantom
acceleration as transient phenomenon with different scalar
potentials or to modify the gravity, couple dark energy with dark
matter or to use particular forms of EoS for dark energy taking into
account some quantum effects (giving rise to the second quantum
gravity era) which may delay/stop the singularity occurrence
\cite{p2}. Inserting Eq.(\ref{3}) into (\ref{1}), we obtain
\begin{equation}\label{4}
\mathcal{I}=\int
d^4x\sqrt{-g}[\frac{f(R,T)}{2\kappa^2}+p_m(a,b)+\frac{\epsilon}{2}
g^{\mu\nu}\partial_{\mu}\phi
\partial_{\nu}\phi-V(\phi)],
\end{equation}
where
\begin{eqnarray}\nonumber
R=\frac{2}{ab^2}\left(\ddot{a}b^2+2ab\ddot{b}
+2b\dot{a}\dot{b}+a\dot{b^2}+a\xi\right),\quad
T=3p_m(a,b)-\rho_m(a,b).
\end{eqnarray}

To evaluate Lagrangian corresponding to the action (\ref{4}) for
configuration space $\mathcal{Q}=\{a,b,R,T,\phi\}$, we use Lagrange
multiplier approach which yields
\begin{eqnarray}\nonumber
\mathcal{L}&=&ab^2[f(R,T)-Rf_R(R,T)+f_T(R,T)(3p_m(a,b)
-\rho_m(a,b)-T)-\frac{\epsilon\dot{\phi}^2}{2}\\\nonumber&+&p_m(a,b)-V(\phi)]
-(4b\dot{a}\dot{b}+2a\dot{b}^2-2a\xi)f_R(R,T)
-(2b^2\dot{a}\dot{R}+4ab\dot{b}\dot{R})\\\label{5}&\times&
f_{RR}(R,T)-(2b^2\dot{a}\dot{T}+4ab\dot{b}\dot{T})f_{RT}(R,T).
\end{eqnarray}
In a dynamical system, the Euler-Lagrange equation, Hamiltonian
($\mathcal{H}$) and conjugate momenta ($p_i$) play a significant
role to determine basic features of the system, defined as
\begin{eqnarray}\nonumber
&&\frac{\partial\mathcal{L}}{\partial
q^i}-\frac{dp_i}{dt}=0,\quad\mathcal{H}=\sum_i\dot{q}^ip_i-\mathcal{L},\quad
p_i=\frac{\partial\mathcal{L}}{\partial \dot{q}^i},
\end{eqnarray}
where $q^i$ shows $n$ coordinates of the system. For Lagrangian
(\ref{5}), the conjugate momenta take the following form
\begin{eqnarray*}
p_a&=&-4b\dot{b}f_R -2b^2(\dot{R}f_{RR}+\dot{T}f_{RT}),\quad
p_{\phi}=-ab^2\epsilon\dot{\phi},\\\nonumber
p_b&=&-4f_R(a\dot{b}+b\dot{a})
-4ab(\dot{R}f_{RR}+\dot{T}f_{RT}),\\\nonumber p_R&=&
-(4ab\dot{b}+2b^2\dot{a})f_{RR},\quad p_T=
-(4ab\dot{b}+2b^2\dot{a})f_{RT}.
\end{eqnarray*}
The dynamical equations of the system are
\begin{eqnarray}\nonumber
&&2f_R(R,T)\left(\frac{\dot{b^2}}{b^2}+\frac{2\ddot{b}}{b}
+\frac{2\xi}{b^2}\right)+f-Rf_R+f_T
(3p_m(a,b)-\rho_m(a,b)-T)\\\nonumber&+&p_m(a,b)-\frac{\epsilon\dot{\phi}^2}{2}
-V(\phi)+a\{f_T(3p_m,_{_a}-\rho_m,_{_a})+p_m,_{_a}\}+4b^{-1}\dot{b}\dot{R}f_{RR}
\\\label{6}&+&4b^{-1}\dot{b}\dot{T}f_{RT}
+2\ddot{R}f_{RR}+2\dot{R}^2f_{RRR}+4\dot{R}\dot{T}f_{RRT}+2\ddot{T}f_{RT}
+2\dot{T}^2f_{RTT}=0,
\\\nonumber&&2f_R\left(\frac{\ddot{a}}{a}+\frac{\dot{a}\dot{b}}{ab}
+\frac{\ddot{b}}{b}\right)+f-Rf_R
+f_T(3p_m(a,b)-\rho_m(a,b)-T)+p_m(a,b)\\\nonumber&-&\frac{\epsilon\dot{\phi}^2}{2}
-V(\phi)
+\frac{b}{2}\{f_T(3p_m,_{_b}-\rho_m,_{_b}))+p_m,_{_b}\}+2(a^{-1}\dot{a}
\dot{R}+\ddot{R})f_{RR}+2\dot{R}^2\\\nonumber&\times&f_{RRR}+2(a^{-1}\dot{a}
\dot{T}+\ddot{T})f_{RT}+2(b^{-1}\dot{b}\dot{R}+2\dot{R}\dot{T}+\dot{T}^2)f_{RRT}
+2b^{-1}\dot{b}\dot{T}f_{RTT}=0,\\\label{7}
\\\nonumber&&f_{RT}(3p_m(a,b)-\rho_m(a,b)-T)=0,\quad f_{TT}(3p_m(a,b)-\rho_m(a,b)-T)=0,
\\\label{8}&&\epsilon\ddot{\phi}+2\epsilon b^{-1}\dot{b}\dot{\phi}
+\epsilon a^{-1}\dot{a}\dot{\phi}-V'(\phi)=0.
\end{eqnarray}
In order to evaluate total energy of the dynamical system, we
formulate Hamiltonian as
\begin{eqnarray}\nonumber
\mathcal{H}&=&2f_R\left(\frac{\dot{b^2}}{b^2}+\frac{2\dot{a}\dot{b}}
{ab}\right)+2\left(\frac{2\dot{b}}{b}+\frac{\dot{a}}{a}\right)
\dot{R}f_{RR}+2\left(\frac{2\dot{b}}{b}+\frac{\dot{a}}{a}\right)
\dot{T}f_{RT}+f-Rf_R\\\label{9}&+&f_T(3p_m(a,b)-\rho_m(a,b)-T)
+p_m(a,b)+\frac{\epsilon\dot{\phi}^2}{2}
-V(\phi)+\frac{2\xi f_R}{b^2}.
\end{eqnarray}
The Hamiltonian constraint $\mathcal{H}=0$ yields total pressure of
the dynamical system.

\section{Noether Symmetry and Conserved Quantities}

The Noether symmetry approach helps to solve complicated non-linear
system of partial differential equations yielding exact solutions at
theoretical grounds of physics and cosmology. Noether theorem states
that if Lagrangian of a dynamical system remains invariant under a
continuous group then group generator leads to the associated
conserved quantity. The conservation of energy and linear momentum
appears for translational invariant Lagrangian in time and position,
respectively whereas the angular momentum is conserved for
rotationally symmetric Lagrangian \cite{13}. In gravitational
theories, the presence of conserved quantities also enhances
physical interpretation of theory but if it does not appreciate the
existence of any conserved quantity, then the theory will be
abandoned due to its non-physical features.

To investigate the existence of Noether symmetry and associated
conserved quantity in non-minimally coupled gravitational theory, we
consider the first order prolongation $K^{[1]}$ of continuous group
defined as
\begin{eqnarray}\label{11}
K^{[1]}=K+(\varphi^j,_t+\varphi^j,_i\dot{q}^i-\vartheta,_t\dot{q}^j
-\vartheta,_i\dot{q}^i\dot{q}^j)\frac{\partial}{\partial\dot{q}^j},
\end{eqnarray}
where cosmic time $t$ is considered to be an affine parameter and
$K$ represents symmetry generator given by
\begin{eqnarray}\label{9}
K&=&\vartheta(t,q^i)\frac{\partial}{\partial
t}+\varphi^j(t,q^i)\frac{\partial}{\partial q^j}.
\end{eqnarray}
Here $\vartheta$ and $\varphi^j$ are unknown coefficients of the
generator. The existence of Noether symmetry is ensured when $K$
follows the invariance condition as
\begin{equation}\label{10}
K^{[1]}\mathcal{L}+(D\vartheta)\mathcal{L}=DB(t,q^i),\quad
D=\frac{\partial}{\partial t}+\dot{q}^i\frac{\partial}{\partial
q^i},
\end{equation}
where $D$ is the total derivative while $B$ represents boundary term
of $K$. When the symmetry generator becomes independent of affine
parameter then boundary term along with first order prolongation
vanishes yielding
\begin{equation}\label{13}
K=\varrho^i(q^i)\frac{\partial}{\partial
q^i}+\left[\frac{d}{dt}(\varrho^i(q^i))\right]\frac{\partial}{\partial\dot{q}^i},
\quad L_K\mathcal{L}=0,
\end{equation}
where $L$ identifies Lie derivative. The symmetries coming from
symmetry generators (\ref{9}) and (\ref{13}) lead to corresponding
conservation law through the first integral defined as
\begin{equation}\label{12}
\Sigma=B-\vartheta\mathcal{L}-(\varphi^j-\dot{q}^j\vartheta)
\frac{\partial\mathcal{L}}{\partial\dot{q}^j},\quad\Sigma=-\eta^j
\frac{\partial\mathcal{L}}{\partial\dot{q}^j}.
\end{equation}
For $Q=\{t,a,b,R,T,\phi\}$, the infinitesimal symmetry generator and
corresponding first order prolongation take the form
\begin{eqnarray}\nonumber
K&=&\tau\frac{\partial}{\partial t}+\alpha\frac{\partial}{\partial
a}+\beta\frac{\partial}{\partial b}+\gamma\frac{\partial}{\partial
R}+\delta\frac{\partial}{\partial
T}+\eta\frac{\partial}{\partial\phi},\quad
K^{[1]}=\tau\frac{\partial}{\partial
t}+\alpha\frac{\partial}{\partial a}+\beta\frac{\partial}{\partial
b}\\\label{14}&+&\gamma\frac{\partial}{\partial
R}+\delta\frac{\partial}{\partial
T}+\dot{\alpha}\frac{\partial}{\partial
\dot{a}}+\dot{\beta}\frac{\partial}{\partial
\dot{b}}+\dot{\gamma}\frac{\partial}{\partial
\dot{R}}+\dot{\delta}\frac{\partial}{\partial\dot{T}}+\dot{\eta}\frac{\partial}{\partial
\dot{\phi}},
\end{eqnarray}
where time derivative of unknown coefficients
$\tau,~\alpha,~\beta,~\gamma,~\delta$ and $\eta$ are
\begin{eqnarray}\label{15}
\dot{\sigma}_{_l}&=&D\sigma_{_l}-\dot{q}^iD\tau,\quad l=1...5,
\end{eqnarray}
Here $\sigma_1,~\sigma_2,~\sigma_3,~\sigma_4$ and $\sigma_5$
correspond to $\alpha,~\beta,~\gamma,~\delta$ and $\eta$,
respectively.

In order to discuss the presence of Noether symmetry generator and
relative conserved quantity of the model (\ref{2}), we insert the
first order prolongation (\ref{11}) alongwith (\ref{9}) in
(\ref{10}), it follows a system of equations given in Appendix
\textbf{A}. From Eq.(\ref{1'}), we have either
$f_R,~f_{RR},~f_{RT}=0$ with
$\tau,_{_a},~\tau,_{_b},~\tau,_{_R},~\tau,_{_T}\neq0$ or vice versa.
For non-trivial solution, we consider second possibility
($\tau,_{_a},~\tau,_{_b},~\tau,_{_R},~\tau,_{_T}=0$) as the first
choice yields trivial solution. We investigate the existence of
symmetry generators, relative conserved quantities for the following
two models \cite{6}
\begin{itemize}
\item $f(R,T)$=$R+2g(T)$,
\item $f(R,T)$=$F(R)+h(R)g(T)$.
\end{itemize}
We also formulate corresponding exact solutions to analyze
cosmological picture of these two models.

\subsection{\textbf{$f(R,T)=R+2g(T)$}}

This model incorporates an indirect non-minimal curvature-matter
coupling and also admits a correspondence with standard cosmological
constant cold dark matter ($\Lambda$CDM) model if it comprises a
trace dependent cosmological constant defined as
\begin{equation}\label{16}
f(R,T)=R+2\Lambda(T)+g(T).
\end{equation}
To evaluate the coefficients of symmetry generator (\ref{9}), we
solve the system (\ref{6'})-(\ref{22'}) via separation of variables
method which gives
\begin{eqnarray}\nonumber
\alpha&=&\alpha_1(t)\alpha_2(a)\alpha_3(b)\alpha_4(R)\alpha_5(T)\alpha_6(\phi),
\quad\delta=\delta_1(t)\delta_2(a)\delta_3(b)\delta_4(R)\delta_5(T)\delta_6(\phi),\\\nonumber
\gamma&=&\gamma_1(t)\gamma_2(a)\gamma_3(b)\gamma_4(R)\gamma_5(T)\gamma_6(\phi),
\quad
\eta=\eta_1(t)\eta_2(a)\eta_3(b)\eta_4(R)\eta_5(T)\eta_6(\phi),\\\nonumber
\beta&=&\beta_1(t)\beta_2(a)\beta_3(b)\beta_4(R)\beta_5(T)\beta_6(\phi),
\quad\tau=\tau_1(t),\\\nonumber
B&=&B_1(t)B_2(a)B_3(b)B_4(R)B_5(T)B_6(\phi).
\end{eqnarray}
For these coefficients, the system (\ref{6'})-(\ref{22'}) yields
\begin{eqnarray}\nonumber
\alpha&=&-2ac_{_1},\quad\beta=c_{_1}b,\quad\gamma=0,\quad\delta=0,\quad\eta=c_{_4},
\\\nonumber B&=&c_{_2}t+c_{_3},\quad\tau=c_{_5},\quad
V(\phi)=c_{_6}\phi+c_{_7},
\\\label{a}p_m(a,b)&=&-\frac{c_{_4}c_{_6}\ln a+2c_{_1}a^{\frac{1}{2}}b}{2c_{_1}}
-\frac{2\xi}{b^2}-\frac{c_{_2}\ln
a}{2c_{_1}ab^2},\\\label{b}\rho_m(a,b)&=&-\frac{3c_{_4}c_{_6}\ln
a+2c_{_1}a^{\frac{1}{2}}b}{2c_{_1}}
-\frac{6\xi}{b^2}-\frac{3c_{_2}\ln a}{2c_{_1}ab^2},
\end{eqnarray}
where $c_i$'s $(i=1...7)$ denotes arbitrary constants. For these
coefficients, we split the symmetry generator and corresponding
first integral into the following form
\begin{eqnarray*}
K_1&=&\frac{\partial}{\partial t},\quad
\Sigma_1=-ab^2\{f-Rf_R+f_T(3p_m-\rho_m-T)+p_m-c_{_6}\phi-c_{_7}\}\\\nonumber&+&2a\xi
f_R-4b\dot{a}\dot{b}f_R-2a\dot{b}^2f_R-\frac{\epsilon\dot{\phi}^2ab^2}{2},
\\\nonumber K_2&=&-2a\frac{\partial}{\partial a}+b\frac{\partial}{\partial
b},\quad\Sigma_2=-4ab\dot{b}f_R+4b^2\dot{a}f_R,\\\nonumber
K_3&=&\frac{\partial}{\partial\phi},\quad\Sigma_3=\epsilon
ab^2\dot{\phi}.
\end{eqnarray*}
For the model (\ref{16}), the system (\ref{6'})-(\ref{22'}) yields
three symmetry generators and associated conserved quantities. In
this case, the symmetry generator $K_1$ leads to energy conservation
while $K_2$ represents scaling symmetry corresponding to
conservation of linear momentum.

Next, we explore the presence of Noether symmetry in the absence of
affine parameter and boundary term of extended symmetry which leads
to establish corresponding conservation law. In this case, the
infinitesimal generator of continuous group for
$\mathcal{Q}=\{a,b,R,T,\phi\}$ turns out to be
\begin{equation}\label{17}
K=\alpha\frac{\partial}{\partial a}+\beta\frac{\partial}{\partial
b}+\gamma\frac{\partial}{\partial R}+\delta\frac{\partial}{\partial
T}+\eta\frac{\partial}{\partial\phi}+\dot{\alpha}\frac{\partial}{\partial
\dot{a}}+\dot{\beta}\frac{\partial}{\partial
\dot{b}}+\dot{\gamma}\frac{\partial}{\partial
\dot{R}}+\dot{\delta}\frac{\partial}{\partial\dot{T}}+\dot{\eta}
\frac{\partial}{\partial\dot{\phi}},
\end{equation}
where $\dot{\alpha}=\dot{q^i}\frac{\partial\alpha}{\partial
q^i},~\dot{\beta}=\dot{q^i}\frac{\partial\beta}{\partial
q^i},~\dot{\gamma}=\dot{q^i}\frac{\partial\gamma}{\partial
q^i},~\dot{\delta}=\dot{q^i}\frac{\partial\delta}{\partial q^i}$ and
$\dot{\eta}=\dot{q^i}\frac{\partial\eta}{\partial q^i}$. Due to the
absence of affine parameter, the separation of variables method
yields
\begin{eqnarray*}
\alpha&=&\alpha_1(a)\alpha_2(b)\alpha_3(R)\alpha_4(T)\alpha_5(\phi),
\quad\beta=\beta_1(a)\beta_2(b)\beta_3(R)\beta_4(T)\beta_5(\phi),\\\nonumber
\gamma&=&\gamma_1(a)\gamma_2(b)\gamma_3(R)\gamma_5(\phi),
\quad\delta=\delta_1(a)\delta_2(b)\delta_3(R)\delta_4(T)\delta_5(\phi),\\\nonumber
\eta&=&\eta_1(a)\eta_2(b)\eta_3(R)\eta_4(T)\eta_5(\phi).
\end{eqnarray*}
In order to explore the consequences of indirect non-minimal
curvature-matter coupling, we evaluate symmetry generators with
corresponding conservation laws for non-existing boundary term. We
also establish cosmological analysis through exact solutions for
both dust and perfect fluid distributions.

\subsubsection*{Dust Case}

Dust fluid investigates matter contents of the universe when the
existence of radiations is not so worthy and the formation of
massive stars is possible only if dust particles interact with
radiations. Here we consider $T_{\mu\nu}=\rho_m u_\mu u_\nu$ and
solve the system for (\ref{17}) via separation of variables which
yields
\begin{eqnarray*}
\alpha&=&-2ac'_{_1},\quad\beta=c'_{_1}b,\quad\gamma=0,\quad\delta=0,\quad\eta=0,
\\\nonumber\rho_m(a,b)&=&
\frac{\xi}{b^2c'_{_2}}+a^{\frac{1}{2}}b,\quad\Lambda(T)=-\frac{g(T)}{2}+c'_{_2}T+c'_{_3},
\end{eqnarray*}
where $c'_{_j}$'s $(j=1...3)$ represents arbitrary constants. The
corresponding symmetry generator and associated conserved quantity
are
\begin{equation*}
K=-2ac'_{_1}\frac{\partial}{\partial
a}+c'_{_1}b\frac{\partial}{\partial
b},\quad\Sigma=4c'_{_1}ab\dot{b}f_R-4c'_{_1}b^2\dot{a}f_R.
\end{equation*}
For dust fluid, there exists only scaling symmetry in the absence of
affine parameter as well as boundary term of extended symmetry and
the model (\ref{16}) reduces to
\begin{equation}\label{18}
f(R,T)=R+2c'_{_2}T+2c'_{_3}.
\end{equation}
For exact solution of equations of motion, we insert density of dust
fluid and model (\ref{18}) in Eqs.(\ref{6}) and (\ref{7}) yielding
\begin{equation*}
a(t)=\frac{(40c'_{_2}t+40c'_{_3})^{\frac{4}{5}}}{16}, \quad
b(t)=\frac{c'_{_1}(40c'_{_2}t+40c'_{_3})^{\frac{2}{5}}}{4}.
\end{equation*}
This leads to expansion of the universe whether it is accelerated or
decelerated. The power-law scale factor $(a(t)=t^\lambda)$
identifies both expansions as for $\lambda>1$, it measures
accelerated expansion while it corresponds to decelerated expansion
for $\lambda<1$. When $\lambda=\frac{1}{2}$ and
$\lambda=\frac{2}{3}$, we have radiation and matter dominated eras
of the universe.

To analyze the behavior of power-law type exact solution, we
construct cosmological analysis through some cosmological parameters
such as Hubble, deceleration, $r-s$ and EoS. These parameters are
useful to study current expansion as well as different eras of the
universe. The Hubble parameter ($H$) determines the rate of
expansion while deceleration parameter ($q$) evaluates the nature of
cosmic expansion whether decelerated ($q>0$), accelerated ($q<0$) or
constant ($q=0$), respectively. In case of anisotropic universe
models, these parameters turn out to be
\begin{eqnarray*}
H=\frac{64c'_{_2}(40c'_{_2}t+40c'_{_3})^{-1}}{3},\quad
q=\frac{7}{8}.
\end{eqnarray*}
The remarkable pair of $r-s$ parameters explores the characteristics
of dark energy candidates by establishing a correspondence between
constructed and standard cosmic models. When the pair lies in
($r,s$)=(1,0) region, this corresponds to standard $\Lambda$CDM
model while the trajectories with $s>0$ and $r<1$ correspond to
quintessence and phantom phases of dark energy. In the present case,
we obtain $r=0$ with $s=-\frac{8}{9}$ indicating that the
constructed model does not correspond to any standard dark energy
universe model. The EoS parameter $(\omega)$ investigates different
cosmic eras such as it identifies radiation and matter dominated
eras for $\omega=\frac{1}{3}$ and $\omega=0$, respectively. This
parameter specifies dark energy era ($\omega=-1$) into quintessence
and phantom phases when $-1<\omega\leq-1/3$ and $\omega<-1$,
respectively. The corresponding effective EoS parameter is
\begin{equation*}
\omega_{eff}=\frac{128c'_{_2}+(40c'_{_2}t+40c'_{_3})^{\frac{4}{5}}
(5c'^2_{_2}c'_{_1}t^2+10c'_{_1}c'_{_2}c'_{_3}t+5c'_{_1}c'^2_{_3})}{128c'_{_2}}
\end{equation*}
The potential and kinetic energies of the scalar field play a
dynamical role to study cosmic expansion. For accelerated expansion,
the field $\phi$ evolves negatively and potential dominates over the
kinetic energy ($\frac{\dot{\phi}^2}{2}<V(\phi)$) whereas negative
potential follows the kinetic energy for decelerated expansion of
the universe ($\frac{\dot{\phi}^2}{2}>-V(\phi)$). Using
Eq.(\ref{8}), we obtain
\begin{eqnarray*}
\phi&=&\int\frac{1}{20\epsilon(c'_{_2}t+c'_{_3})}\left(\left(-\epsilon
c'_{_2}\left(25c'^2_{_2}(40c'_{_2}t+40c'_{_3})^{\frac{4}{5}}c'_{_1}t^2+50
(40c'_{_2}t+40c'_{_3})^{\frac{4}{5}}\right.\right.\right.\\\nonumber&\times&
\left.\left.\left.c'_{_2}c'_{_1}c'_{_3}t+25(40c'_{_2}t
+40c'_{_3})^{\frac{4}{5}}c'_{_1}c'^2_{_3}+896c'_{_2}\right)
\right)^{\frac{1}{2}}\right),\\\nonumber
V(\phi)&=&\frac{1}{800\left(c'^2_{_2}t^2
+2c'_{_2}c'_{_3}t+c'^2_{_3}\right)}\left[25c'_{_1}\left(5c'^3_{_2}t^2
+5c'_{_2}c'^2_{_3}+10c'^2_{_2}c'_{_3}t\right)(40c'_{_2}t
\right.\\\nonumber&+&\left.40c'_{_3})^{\frac{4}{5}}+8c'_{_2}c'_{_3}\left(-200c'_{_2}
t^2+400c'_{_3}t\right)-8\left(48c'^2_{_2}+200c'^3_{_3}\right)\right].
\end{eqnarray*}
\begin{figure}\epsfig{file=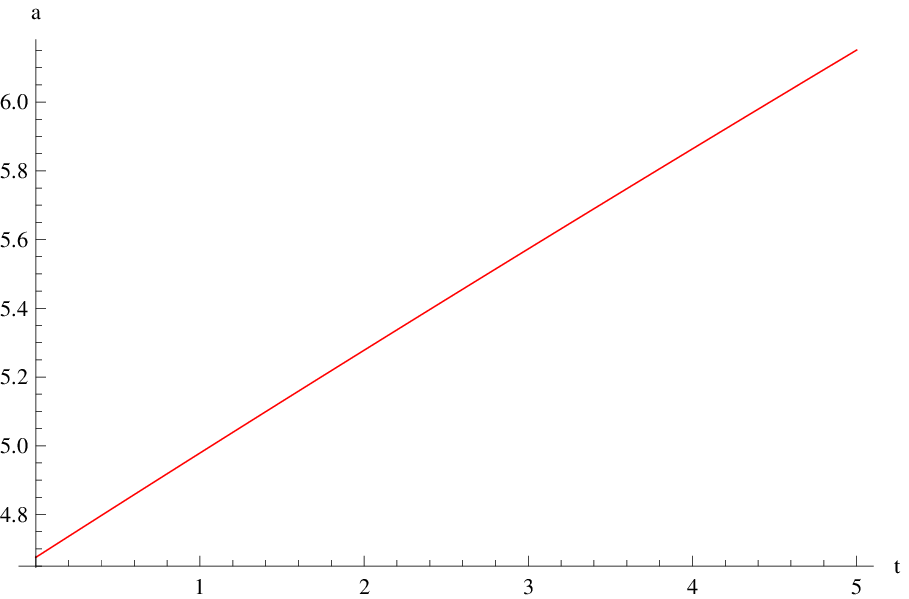,
width=0.45\linewidth}\epsfig{file=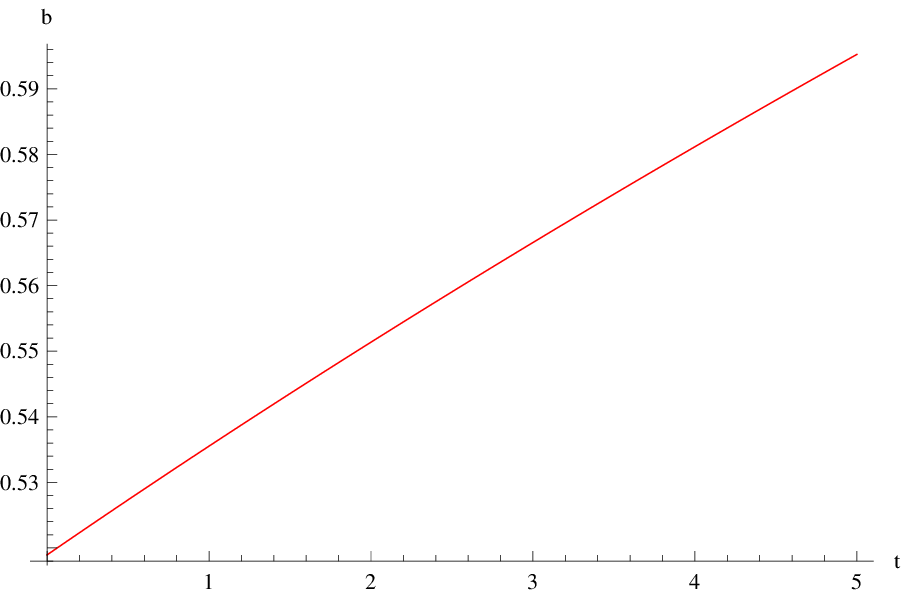,
width=0.45\linewidth}\caption{Plots of scale factors $a(t)$ (left)
and $b(t)$ (right) versus cosmic time $t$ for $c'_{_1}=0.24$,
$c'_{_2}=0.45$ and $c'_{_3}=5.5$.}
\end{figure}
\begin{figure}\epsfig{file=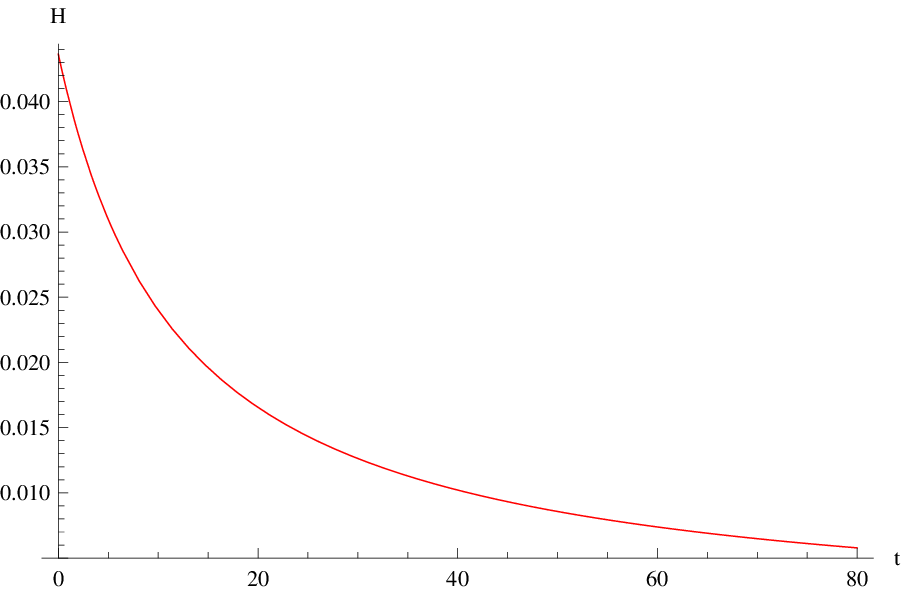,
width=0.45\linewidth}\epsfig{file=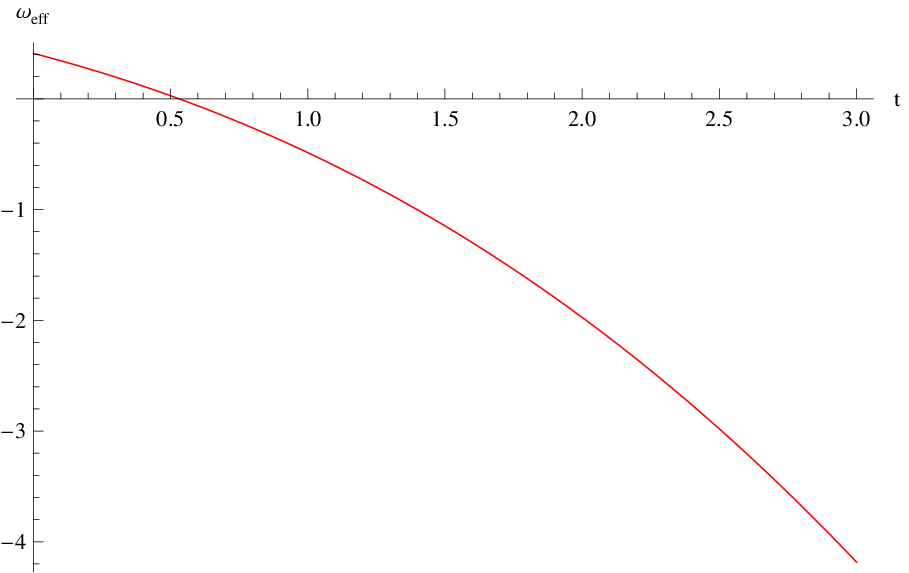,
width=0.45\linewidth}\caption{Plots of Hubble $H(t)$ (left) and EoS
parameters $\omega_{eff}$ (right) versus cosmic time $t$.}
\end{figure}

Figure \textbf{1} shows the graphical analysis of the scale factors
for the dust case. The scale factor $a(t)$ indicates large cosmic
expansion in $x$-direction but $b(t)$ represents that the universe
is expanding very slowly in $y$ and $z$-directions. Figure
\textbf{2} (left plot) indicates that Hubble parameter is decreasing
with the passage of time. In the right plot of Figure \textbf{2},
the effective EoS parameter identifies that initially, the universe
appreciates radiation dominated era and after sometime, it
corresponds to dark energy era by crossing matter dominated phase.

Figures \textbf{3} and \textbf{4} analyze the behavior of scalar
field and cosmic expansion via phantom and quintessence models. The
left plot of Figure \textbf{3} shows that the scalar field is
positive initially yielding decelerated expansion but gradually, it
starts increasing negatively which describes accelerated expansion.
In case of quintessence model, the scalar field grows from negative
to positive indicating decelerated expansion of the universe. The
right plots of \textbf{3} and \textbf{4} satisfy
$\frac{\dot{\phi}^2}{2}<V(\phi)$ and
$\frac{\dot{\phi}^2}{2}>-V(\phi)$ implying that phantom model yields
accelerated expansion while quintessence model corresponds to
decelerated expansion.
\begin{figure}\epsfig{file=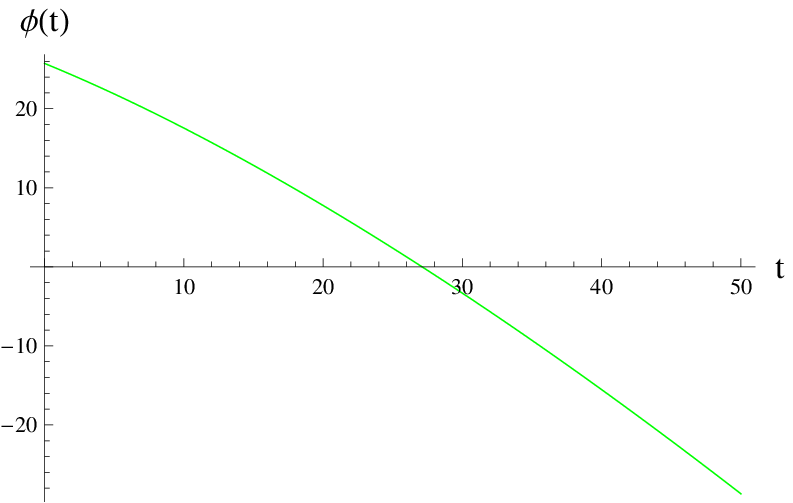,
width=0.45\linewidth}\epsfig{file=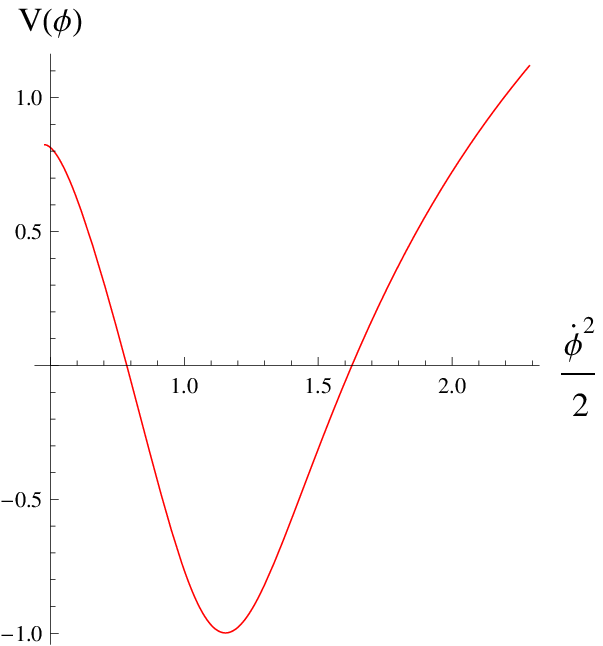,
width=0.3\linewidth}\caption{Plots of scalar field $\phi(t)$ (left)
versus cosmic time $t$ and potential energy $V(\phi)$ versus kinetic
energy $\frac{\dot{\phi}^2}{2}$ (right) for $\epsilon=-1$.}
\end{figure}
\begin{figure}\epsfig{file=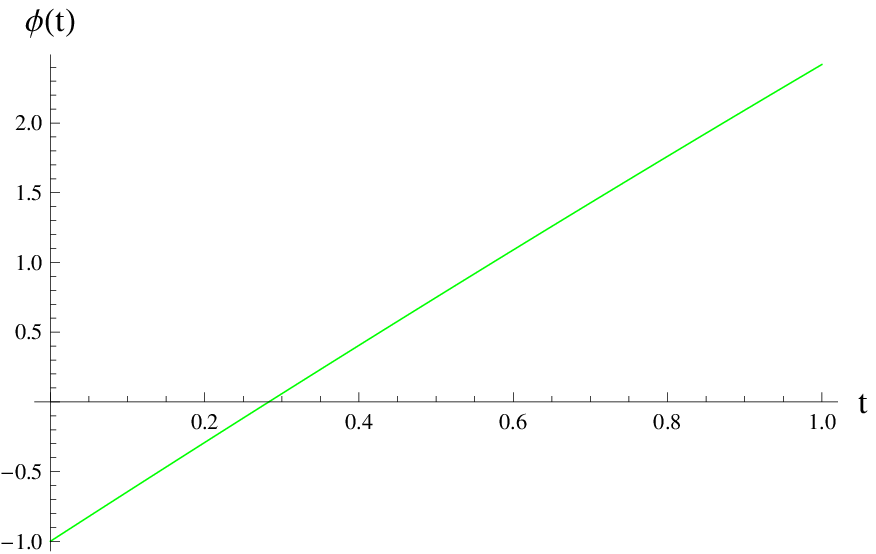,
width=0.45\linewidth}\epsfig{file=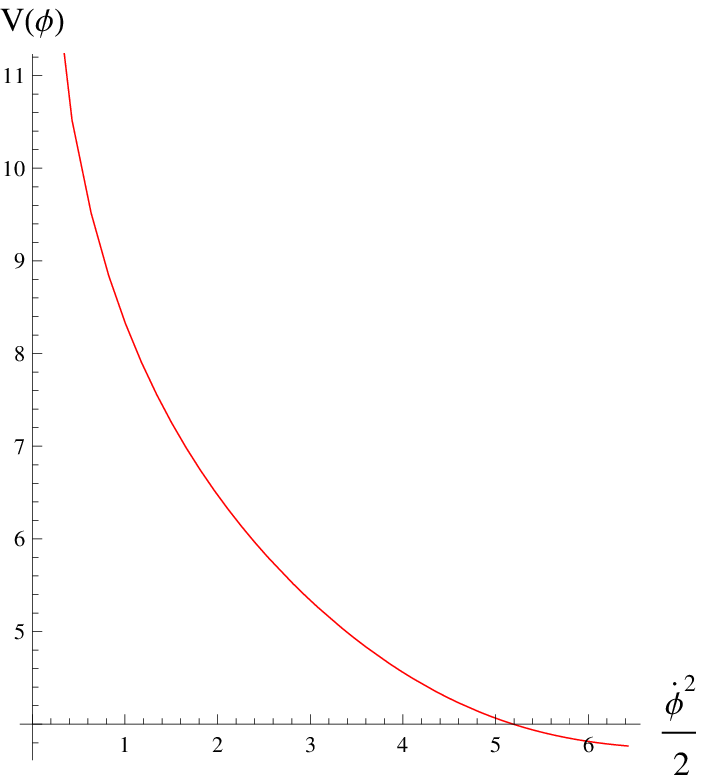,
width=0.3\linewidth}\caption{Plots of scalar field $\phi(t)$ (left)
versus cosmic time $t$ and potential energy $V(\phi)$ versus kinetic
energy $\frac{\dot{\phi}^2}{2}$ (right) for $\epsilon=1$.}
\end{figure}

To analyze a big-rip free model, the key point is that if EoS
parameter rapidly approaches to -1 and Hubble rate tends to be
constant (asymptotically de Sitter universe), then it is possible to
have a model in which the time required for singularity is infinite,
i.e., the singularity effectively does not occur \cite{p3}. The
occurrence of maximum potential of a phantom scalar field is another
evident fact to avoid this singularity \cite{p4}. The graphical
behavior of EoS parameter represents that $\omega_{eff}$ rapidly
approaches to -1 and Hubble rate is decreasing but potential is not
maximum. We may avoid the big-rip singularity in the present case if
we choose $c_2'$ to be negatively large that yields asymptotic
behavior of Hubble rate.

\subsubsection*{Non-Dust Case}

At large scales, perfect fluid successfully illustrates cosmic
matter distribution in the presence of radiations. In the absence of
boundary term and affine parameter, the coefficients of symmetry
generator (\ref{17}) corresponding to $a,b,R,T,\phi$ remain the same
as in the presence of boundary term of extended symmetry. Thus,
generator of Noether symmetry and associated first integrals reduce
to
\begin{eqnarray*}
K&=&-2ac_{_1}\frac{\partial}{\partial
a}+c_{_1}b\frac{\partial}{\partial
b}+c_{_2}\frac{\partial}{\partial\phi},\\\nonumber\Sigma&=&
-4c_{_1}ab\dot{b}f_R+4c_{_1}b^2\dot{a}f_R+\epsilon
c_{_2}ab^2\dot{\phi}.
\end{eqnarray*}
In order to formulate exact solution of dynamical equations for
perfect fluid distribution, we insert Eqs.(\ref{a}) and (\ref{b})
into (\ref{6}) and (\ref{7}) yielding
\begin{eqnarray*}
a(t)&=&\frac{\left(\frac{5}{c_{_9}}\right)^{\frac{2}{5}}(c_{_2}\sin(c_{_{10}}t)
+c_{_3}\cos(c_{_{10}}t))^{\frac{4}{5}}}{5^{\frac{4}{5}}},\\\nonumber
b(t)&=&\frac{c_{_4}\left(\frac{5}{c_{_9}}\right)^{\frac{1}{5}}(c_{_2}\sin(c_{_{10}}t)
+c_{_3}\cos(c_{_{10}}t))^{\frac{2}{5}}}{5^{\frac{2}{5}}}.
\end{eqnarray*}
This describes oscillatory solution of $f(R,T)$ model admitting
indirect non-minimal curvature-matter coupling. To study the
cosmological behavior of this solution, we consider cosmological
parameters as follows
\begin{eqnarray*}\nonumber
H&=&\frac{8c_{_{10}}(c_{_2}\sin(c_{_{10}}t)
+c_{_3}\cos(c_{_{10}}t))}{15(c_{_2}\sin(c_{_{10}}t)
+c_{_3}\cos(c_{_{10}}t))},\\\nonumber
q&=&\frac{-8c^2_{_2}\cos^2(c_{_{10}}t)+7c^3_{_3}+8c^2_{_3}
\cos^2(c_{_{10}}t)+15c^2_{_2}+16c_{_2}c_{_3}\cos(c_{_{10}}t)
\sin(c_{_{10}}t)}{8(c_{_2}\sin(c_{_{10}}t)
+c_{_3}\cos(c_{_{10}}t))^2},\\\nonumber
s&=&(-45((4c^4_{_2}-4c^4_{_3})\cos^2(c_{_{10}}t)
-c^4_{_3}-6c^2_{_2}c_{_3}-5c^4_{_2}-(8c^3_{_2}c_{_3}+8c_{_2}
c^3_{_3})\\\nonumber&\times&\cos(c_{_{10}}t)\sin(c_{_{10}}t)))
/256((c^4_{_2}+c^4_{_3}-6c^2_{_2}c^2_{_3})\cos^4(c_{_{10}}t)
+(6c^2_{_2}c^2_{_3}-2c^4_{_3})\\\nonumber&\times&\cos^2(c_{_{10}}t)
+(-4c^3_{_2}c_{_3}+4c_{_2}c^3_{_3})\sin(c_{_{10}}t)
\cos^3(c_{_{10}}t)-4c_{_2}c^3_{_3}\cos(c_{_{10}}t)\\\nonumber&\times&
\sin(c_{_{10}}t)+c^4_{_3}),\\\nonumber
\omega_{eff}&=&\frac{\chi(3p_m-\rho_m)+p_m
-\frac{\epsilon\dot{\phi}^2}{2}-V(\phi)+\frac{2\xi}{b^2}
+a(3p_m,_{_a}-\rho_m,_{_a})+p_m,_{_a}}{\chi(3p_m-\rho_m)+p_m
+\frac{\epsilon\dot{\phi}^2}{2}-V(\phi)+\frac{2\xi}{b^2}}.
\end{eqnarray*}
\begin{figure}\epsfig{file=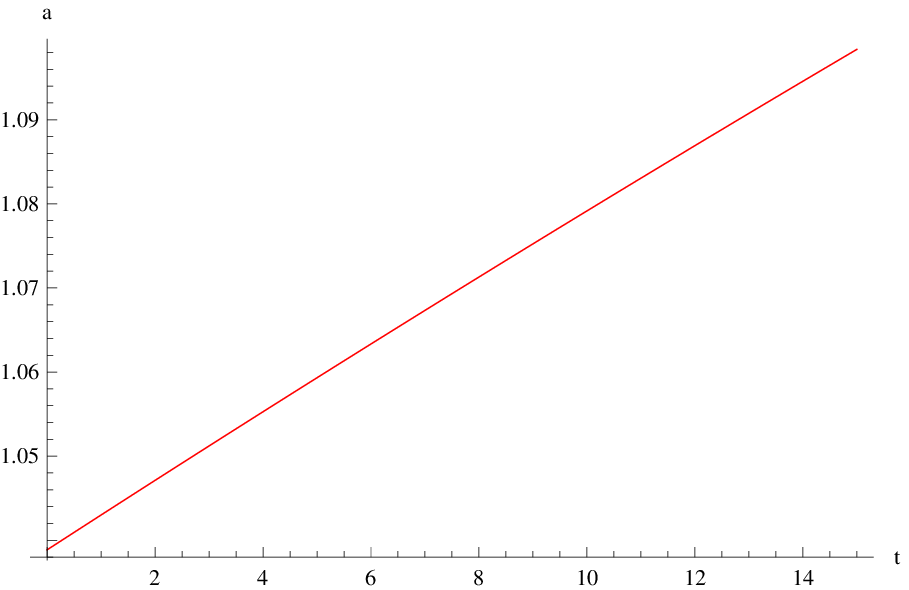,
width=0.5\linewidth}\epsfig{file=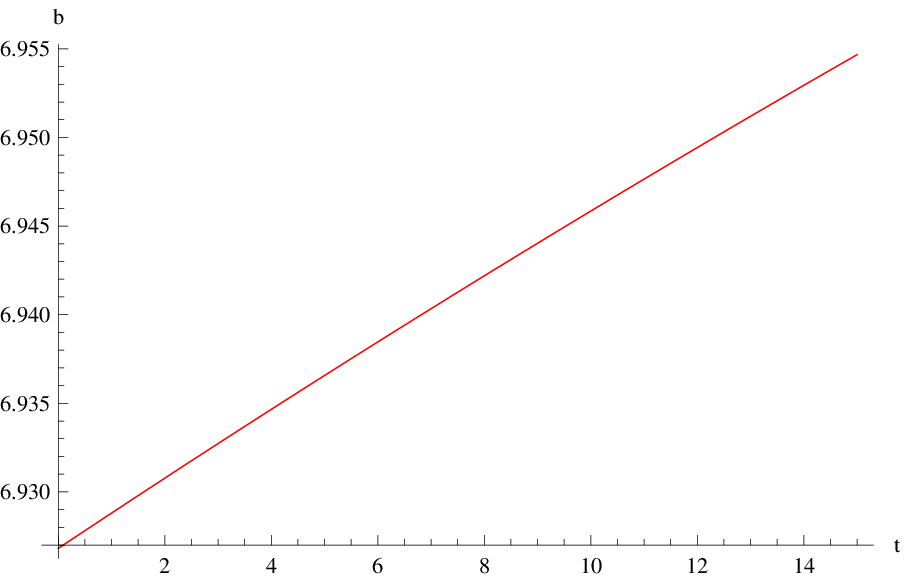,
width=0.5\linewidth}\caption{Plots of scale factor $a(t)$ (left) and
$b(t)$ (right) versus cosmic time $t$ for $c_2=c_3=c_9=5.5$ and
$c_{10}=0.005$.}
\end{figure}
\begin{figure}\epsfig{file=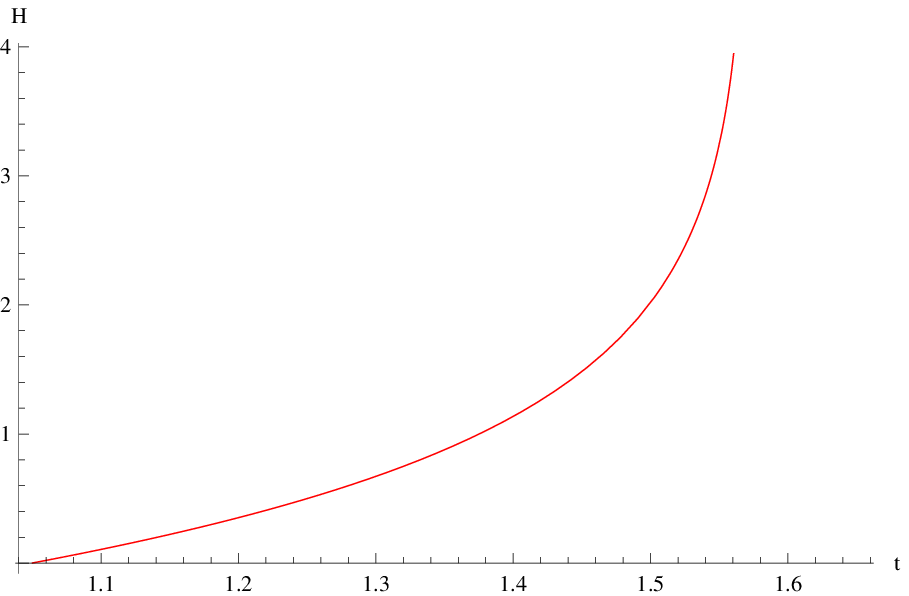,
width=0.5\linewidth}\epsfig{file=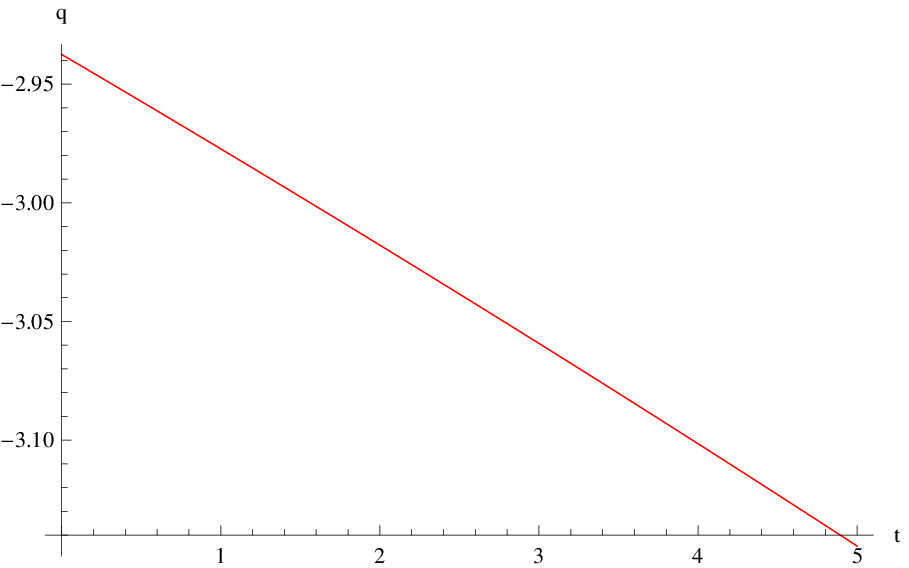,
width=0.5\linewidth}\caption{Plots of $H(t)$ (left) and $q(t)$
(right) versus cosmic time $t$.}
\end{figure}
The scalar field as well as corresponding kinetic and potential
energies identify the early as well as current cosmic expansion and
also characterize decelerated expansion of the universe when kinetic
energy dominates negative potential. In this case, Eq.(\ref{8})
yields
\begin{equation*}
\phi=\int\frac{\epsilon
c_{_4}-\frac{5c_{_6}c^2_{_2}\cos(2c_{_{10}}t)
\left(-2_{2}F_{1}\left[\frac{3}{10},\frac{1}{2},\frac{13}{10},\sin\left[\frac{\pi
}{4}+c_{_{10}}t\right]^2\right]+\sqrt{2-2\sin[2c_{_{10}}t]}\right)}{16
c_{_{10}}\sqrt{\cos\left[\frac{\pi}{4}+c_{_{10}}t\right]^2}(c_{_2}
(\cos[c_{_{10}}t]+\sin[c_{_{10}}t]))^{2/5}}dt}{\epsilon(c_{_2}
\cos[c_{_{10}}t]+c_{_2}\sin[c_{_{10}}t])^{8/5}},
\end{equation*}
where $_{2}F_{1}$ represents hypergeometric function.
\begin{figure}\centering\epsfig{file=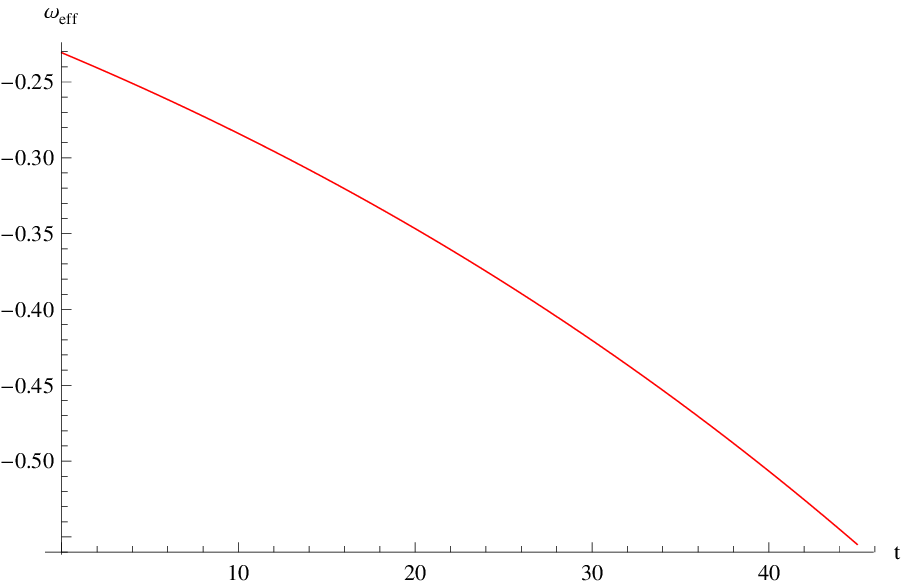,
width=0.5\linewidth}\epsfig{file=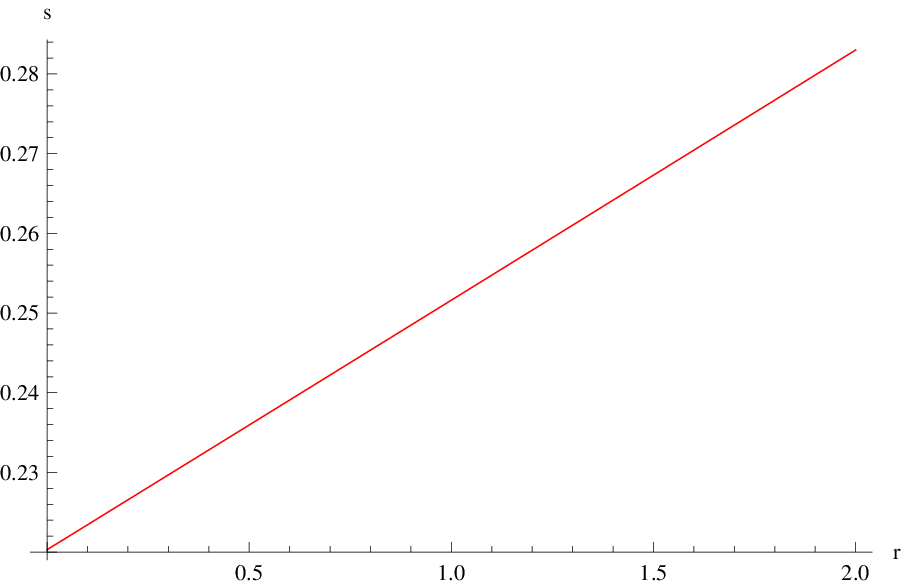,
width=0.5\linewidth}\caption{Plot of $\omega_{eff}$ and $r-s$
parameters versus cosmic time $t$ for $c_2=c_3=5.5$ and
$c_{10}=0.005$.}
\end{figure}
\begin{figure}\epsfig{file=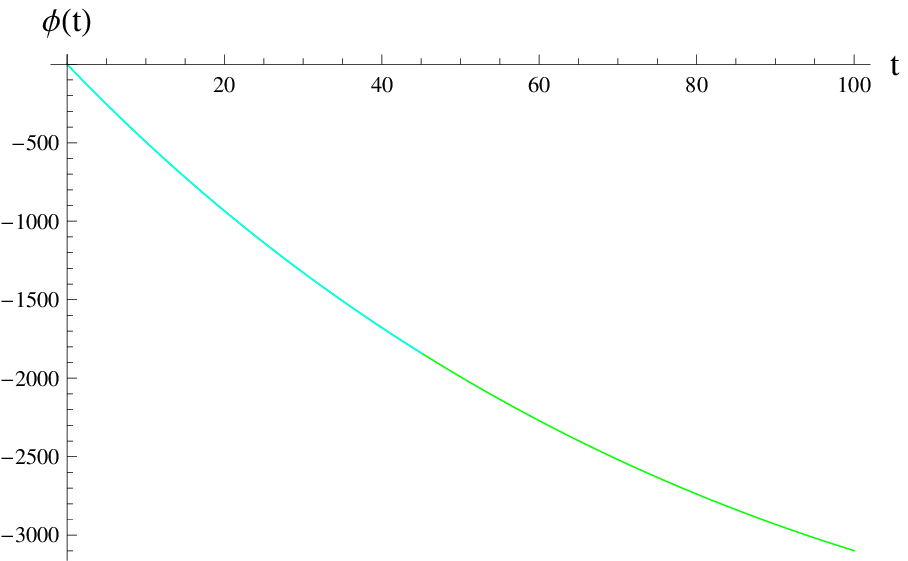,
width=0.5\linewidth}\epsfig{file=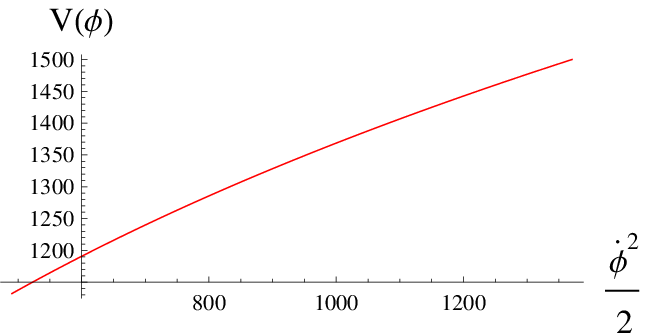,
width=0.5\linewidth}\caption{Plots of scalar field $\phi(t)$ (left)
versus cosmic time $t$ and potential energy $V(\phi)$ versus kinetic
energy $\frac{\dot{\phi}^2}{2}$ (right) for $c_2=5.5$,
$c_4=-10^{3}$, $c_6=0.5$ and $c_{10}=0.005$.}
\end{figure}

In Figure \textbf{5}, the right plot shows that the universe
experiences immense amount of expansion in $y$ and $z$-directions
whereas the left plot yields a small amount of expansion in
$x$-direction. Figure \textbf{6} provides information about
increasing rate of expansion through Hubble parameter while
negatively increasing deceleration parameter assures accelerated
cosmic expansion. The left plot of Figure \textbf{7} characterizes
quintessence phase of dark energy era while the right plot
identifies the $r-s$ parameters trajectories in quintessence and
phantom phases as $s>0$ when $r<1$. Both plots of Figure \textbf{8}
verify the current cosmic expansion for quintessence as well as
phantom models as $\phi$ continuously increasing negatively and
potential energy of the field is dominating over kinetic energy. The
graphical interpretation of EoS parameter yields $\omega_{eff}<-1$
which is not a sufficient condition for the existence of singularity
as potential turns out to be maximum with the passage of time. Thus,
we may avoid big-rip singularity if Hubble rate decreases
asymptotically in the presence of minimal coupling of $f(R,T)$
gravity with scalar field.

\subsection{\textbf{$f(R,T)$=$F(R)+h(R)g(T)$}}

To analyze the effect of direct non-minimal curvature-matter
coupling, we consider this model and evaluate symmetry generators as
well as associated conservation laws. Inserting the model in
Eqs.(\ref{a'})-(\ref{21'}), (\ref{16'}), (\ref{17'}) and (\ref{19'})
and using separation of variables approach, we obtain
\begin{eqnarray*}
\beta&=&-\frac{b\alpha}{2a}+\phi Y_1(t,a,b)+ Y_2(t,a,b),\\\nonumber
F(R)&=&\frac{\epsilon}{4d_3}\left(-d_3Y_{12}(R)+d_2Y_9(R)\right)+d_5R+d_6,
\\\nonumber h(R)&=&-\frac{\epsilon}{4d_3}\left(-d_3Y_{9}(R),_{_R}+d_1R\right)+d_4,
\quad g(T)=d_2+d_3Y_{10}(T),
\\\nonumber\eta&=&\frac{1}{b}\left[Y_1(t,a,b)(Y_{10}(T)(d_1+Y_{9}(R),_{_R})
-\phi^2+Y_{12}(R),_{_R})+b\phi\tau,_{_t}-2\phi\right.\\\nonumber&\times&\left.
Y_2(t,a,b)+b Y_{14}(t,a,b)\right],
\end{eqnarray*}
where $d_i$'s ($i=1...7$) denote constants. We substitute these
values in Eqs.(\ref{6'}), (\ref{4'}) and (\ref{5'}) which yield
\begin{eqnarray*}
\tau&=&\int-\frac{Y_{23}(t)}{\epsilon}dt+d_8t+d_9, \quad
B=\frac{1}{6d_4}\left[6ab(Y_{19}(T)d_1
+d_4\epsilon\phi^2\right.\\\nonumber&+&\left.Y_{19}(T)
d_4e^{-R})Y_2(t,a,b),_{_t}+6ab\phi
(\frac{1}{3}d_4\epsilon\phi^2+Y_{19}(T)d_4e^{-R}
+Y_{19}(T)\right.\\\nonumber&\times&\left.d_1)Y_{16}
(t,b),_{_t}+3d_4(2Y_{22}(t,a,b)+2\phi
Y_{21}(t,a,b),_{_t}+ab^2\phi^2Y_{23}(t),_{_t})\right],
\\\nonumber Y_1(t,a,b)&=&Y_{16}(t,b)+Y_{15}(a,b)\quad Y_{10}(T)=
-\frac{Y_{19}(T)d_3+\epsilon d_2d_4}{\epsilon d_3d_4},\\\nonumber
Y_{12}(R)&=&-\frac{d_2d_4e^{-R}}{d_3}+d_6R+d_7,\quad Y_{9}(R)=
-e^{-R}d_4-2d_1R+d_2,\\\nonumber Y_{14}(t,a,b)&=&-\frac{Y_{21}(t, a,
b)}{b^2a\epsilon}-\frac{bad_2\epsilon d_1Y_{16}(t, b)+d_6\epsilon
bad_3Y_{16}(t, b)}{\epsilon b^2ad_3}+Y_{24}(b, a).
\end{eqnarray*}
To evaluate remaining unknown functions, we insert the above
functions into $\beta,~\eta,~F,~g,~h$ and solve
Eqs.(\ref{10'})-(\ref{1'}) with (\ref{11'})-(\ref{3'}) and
(\ref{12'})-(\ref{14'}) leading to
\begin{eqnarray*}
Y_{21}(t, a, b)&=&Y_{26}(a, b),\quad Y_{22}(t, a, b)=d_{10}t, \quad
Y_{16}(t, b)=-d_{12}b,\\\nonumber Y_{15}(a,b)&=&d_{12}b,\quad
Y_{24}(b,a)=0,\quad Y_{2}(t,a,b)=d_9 b,\\\nonumber
Y_{23}(t)&=&\epsilon(-2
d_9+e^{-R}d_4d_3d_{11}e^{R}+d_8),\quad\delta=0,\quad\gamma=
\frac{d_{11}e^{R}T}{d_13}\\\nonumber&\times&(e^{-R}d_4Td_{13}d_3-d_1d_{13}d_3T+(2((-2d_5
+\frac{1}{2}\epsilon d_6)d_3+d_1d_2\epsilon))d_4).
\end{eqnarray*}
Using these solutions in Eq.(\ref{22'}) with $d_{11}$=0 and
$d_6=\frac{d_2d_1}{d_3}$, it follows that
\begin{eqnarray*}
\tau&=&3d_9,\quad\alpha=d_{10}a,\quad\beta=b(d_9-\frac{d_{10}}{2}),
\quad\delta=0,\quad\gamma=0,\\\nonumber
B&=&d_{10}t,\quad\eta=-\frac{d_1}{\epsilon}+2d_{12}d_6,\quad F(R)=
d_6+d_5R-\frac{3d_6\epsilon R}{4},\\\nonumber
h(R)&=&d_4-\frac{\epsilon}{4d_3}(d_4e^{-R}+d_1R-2d_1),\quad
g(T)=d_2-\frac{d_2d_4\epsilon-d_3d_{13}T}{d_4\epsilon}.
\end{eqnarray*}
Inserting $F,~h$ and $g$, the $f(R,T)$ model becomes
\begin{equation*}
f(R,T)=-\frac{3\epsilon
d_6R}{4}+d_5R+d_6+(d_4-\frac{\epsilon}{4d_3}(d_4e^{-R}+d_1R-2d_1))
(\frac{d_3d_{13}T}{d_4\epsilon}).
\end{equation*}
Thus, the constructed model also experiences a direct coupling
between curvature and matter parts. In this case, the symmetry
generators and associated conserved quantities are
\begin{eqnarray*}
K_1&=&3\frac{\partial}{\partial t}+b\frac{\partial}{\partial b},
\quad\Sigma_1=\frac{1}{4d_3\epsilon}(-4ab^2\epsilon^2d_3\dot{\phi}^2
+4d_{10}d_3\epsilon t+3tab^2d_1RTd_3\epsilon\\\nonumber&-&
9tab^2d_1Rp_md_3\epsilon+3tab^2d_1R\rho_m
d_3\epsilon-12td_1T\dot{a}\dot{b}bd_3\epsilon
-4b^2d_4T\\\nonumber&\times&\dot{a}e^{-R}d_3\epsilon-4b^2a\dot{T}
d_4e^{-R}d_3\epsilon-9tab^2d_4p_me^{-R}d_3\epsilon
+3tab^2d_4\rho_m\\\nonumber&\times&e^{-R}d_3\epsilon+
6td_4Ta\dot{b}^2e^{-R}d_3\epsilon+6td_4Taqe^{-R}d_3\epsilon
-4bd_4Ta\dot{b}e^{-R}d_3\epsilon\\\nonumber&+&4b^2a\dot{R}d_4Te^{-R}
d_3\epsilon+4b^2a\dot{T}d_1d_3\epsilon-12tab^2
d_2d_1\epsilon-12tab^2p_m\\\nonumber&\times&d_3\epsilon+12tab^2
V(\phi)d_3\epsilon-24td_5a\dot{b}d_3\epsilon
-24td_5aqd_3\epsilon+16bd_5a\dot{b}
\\\nonumber&\times&d_3\epsilon
+36tab^2d_3^2d_4p_m-12tab^2d_3^2d_4\rho_m
+18t\epsilon^2d_2d_1a\dot{b}^2+18t\epsilon^2
\\\nonumber&\times&d_2d_1aq
-12b\epsilon^2d_2d_1a\dot{b}+4b^2d_1T\dot{a}d_3\epsilon
-12b^2\epsilon^2d_2d_1\dot{a}+16b^2d_5\\\nonumber&\times&
\dot{a}d_3\epsilon
-3tab^2Rd_4Te^{-R}d_3\epsilon+12td_4T\dot{a}\dot{b}be^{-R}
d_3\epsilon+36t\epsilon^2d_2d_1\dot{a}\dot{b}b\\\nonumber&+&
4bd_1Ta\dot{b} d_3\epsilon+6tab^2e\dot{\phi}^2d_3\epsilon
+18tab^2d_1p_md_3\epsilon-6tab^2d_1\rho_md_3\epsilon
\\\nonumber&-&48td_5\dot{a}\dot{b}bd_3\epsilon-6td_1Ta\dot{b}^2
d_3\epsilon-6td_1Taqd_3\epsilon),\\\nonumber
K_2&=&a\frac{\partial}{\partial
a}-\frac{b}{2}\frac{\partial}{\partial
b},\quad\Sigma_2=-\frac{b}{2d_3}(-a\dot{b}d_1Td_3+3a\dot{b}\epsilon
d_2d_1+a\dot{b}d_4Te^{-R}d_3\\\nonumber&-&4a\dot{b}d_5d_3+bd_1T\dot{a}d_3
+4bd_5\dot{a}d_3-3b\epsilon d_2d_1\dot{a}-bd_4T\dot{a}e^{-R}d_3),
\\\nonumber K_3&=&-\frac{1}{\epsilon}\frac{\partial}{\partial
\phi},\quad\Sigma_3=ab^2\dot{\phi},\quad
K_4=2d_{12}\frac{\partial}{\partial
\phi},\quad\Sigma_4=2d_{12}ab^2\epsilon\dot{\phi}.
\end{eqnarray*}
We see that scaling symmetry appears through generator $K_2$ with
the first integral $\Sigma_2$ leading to conserved linear momentum.

Now we investigate the existence of Noether symmetry in the absence
of affine parameter and boundary term of the extended symmetry and
also study the effect of direct curvature-matter coupling on
conservation laws. For this purpose, we solve Eqs.(\ref{10'}),
(\ref{15'}), (\ref{5'}) and (\ref{11'})-(\ref{14'}) which give
\begin{eqnarray*}
\delta&=& -\frac{a}{2Y_9(T),_{_T}}(\frac{1}{3}Y_4(a, b),_{
_a}\phi^3+2Y_4(a, b),_{ _a}Y_9(T)\phi+2Y_4(a, b),_{
_a}Y_8(b)\phi\\\nonumber&+&\phi^2Y_5(a, b),_{ _a}+2Y_7(a,
b),_{_a}\phi)+Y_{12}(a,R,T,b),\quad
F(R)=k_4R+k_5,\\\nonumber\beta&=&-\frac{b}{2a}(Y_{10}(a,
R,T,b)+aY_5(a, b)),\quad g(T)=k_1+Y_9(T)k_2\\\nonumber\eta&=&
\frac{1}{2}(\phi^2+2Y_9(T)+2Y_8(b))Y_4(a, b)+Y_5(a, b)\phi+Y_7(b,
a),\\\nonumber h(R)&=&\frac{\epsilon
R}{2(k_2+k_3)},\quad\gamma=Y_{11}(a,b,R,T),\\\nonumber\alpha&=&-Y_4(a,
b)a\phi+Y_{10}(a,b,R,T),
\end{eqnarray*}
where $k_l$'s ($l=1...5$) are arbitrary constants. Inserting these
solutions into the remaining equations of the system, we obtain
\begin{eqnarray*}
V(\phi)&=&k_{10}\phi+k_{11},\quad Y_{10}(a,R,T,b)=k_8a,\quad Y_4(a,
b)= 0,\\\nonumber Y_{12}(a, R, T, b) &=&
-\frac{k_8}{2k_2}((\epsilon(k_6T+k_7)+2k_4)k_2+\epsilon k_1),\quad
Y_5(a, b)=-\frac{k_6k_8\epsilon}{2},\\\nonumber Y_7(b, a)&=&
k_9,\quad Y_9(T) = k_6T+k_7, \\\nonumber p_m&=&\frac{
2k_9k_{10}}{\epsilon
k_8k_6}-k_5+k_{11}+\frac{2k_2k_4k_3}{\epsilon}+a^{-\frac{k_6\epsilon}{2}}\epsilon
k_6ba^{\frac{1}{2}-\frac{\epsilon k_6}{4}},\\\nonumber
\rho_m&=&\frac{k_7}{k_6}+\frac{6k_2k_4k_3}{\epsilon}
-3k_5+3k_{11}+\frac{2k_4}{k_6\epsilon}+\frac{6k_9k_{10}}
{k_8k_6\epsilon}+\frac{k_1}{k_2k_6}\\\nonumber&+&a^{-\frac{k_6\epsilon}{2}}\epsilon
k_6ba^{\frac{1}{2}-\frac{\epsilon k_6}{4}}.
\end{eqnarray*}
The corresponding Noether symmetry generator with associated first
integral take the form
\begin{eqnarray*}
K_1&=&a\frac{\partial}{\partial
a}-\frac{b}{2}\left(1-\frac{k_6\epsilon}{2}\right)\frac{\partial}{\partial
b}+R\frac{\partial}{\partial
R}-(k_1\epsilon+k_2(\epsilon(k_6T+k_7)+2k_4))\\\nonumber&\times&
\frac{1}{2k_2}\frac{\partial}{\partial
T}-\frac{k_6\epsilon\phi}{2}\frac{\partial}{\partial\phi},\quad
\Sigma_1=ab\dot{b}\epsilon k_6T-\frac{b^2\epsilon
k_1\dot{a}}{k_2}-b^2\epsilon k_6T\dot{a}-ba\epsilon
k_6k_4\dot{b}\\\nonumber&-&\frac{ba\epsilon^2k_6^2T\dot{b}}{2}
-\frac{ba\epsilon^2k_6k_7\dot{b}}{2}+\frac{ab\dot{b}\epsilon
k_1}{k_2}+ab\dot{b}\epsilon k_7
-\frac{ba\epsilon^2k_6k_1\dot{b}}{2k_2}
-\frac{ab^2\epsilon^2\dot{\phi}k_6\phi}{2}\\\nonumber&+&2ab\dot{b}k_4
-2b^2k_4\dot{a}-b^2\epsilon
k_7\dot{a}+\frac{b^2a\epsilon^2k_6^2\dot{T}}{2} ,\\\nonumber
K_2&=&\frac{\partial}{\partial
\phi},\quad\Sigma_2=ab^2\epsilon\dot{\phi}k_9.
\end{eqnarray*}
Here the symmetry generator $K_1$ yields scaling symmetry.

\section{Final Remarks}

In this paper, we have analyzed the existence of Noether symmetry in
a non-minimally coupled $f(R,T)$ gravity interacting with scalar
field model for anisotropic homogeneous universe models like BI,
BIII and KS models. Using Noether symmetry approach, we have found
conserved quantities associated with symmetry generators and studied
the contribution of direct as well as indirect curvature-matter
coupling through two $f(R,T)$ models. We have also formulated exact
solutions for dust and perfect fluid distributions whose
cosmological analysis is discussed through cosmological parameters.

For $f(R,T)$ model admitting indirect curvature-matter coupling, we
have found three symmetry generators in the presence of affine
parameter and boundary term. The first generator of translational
symmetry in time yields energy conservation law whereas the second
generates scaling symmetry. For the second model, we have formulated
four conserved quantities associated with symmetry generators but
only one generator provides scaling symmetry leading to the
conservation of linear momentum. In the absence of boundary term of
extended symmetry and affine parameter, the symmetry generator of
first model assures the existence of scaling symmetry for dust as
well as perfect fluid while we have found two symmetry generators
for the second model.

For the first model, we have evaluated exact solutions without
considering boundary term. For dust distribution, we have found
power-law solution. The graphical analysis of scale factors and
cosmological parameters lead to decelerating phase of the universe.
The positively increasing scalar field and dominating kinetic energy
over potential energy ensure the decelerating behavior of cosmos for
quintessence model. In case of phantom model, the scalar field rolls
down positively and tends to increase negatively while kinetic
energy dominates over potential energy for $t\in[0.8,1.6]$. The
graphical behavior of effective EoS parameter reveals that the
universe experiences a phase transition from radiation dominated era
to dark energy era by crossing matter dominated phase. For perfect
fluid, we have determined an oscillatory solution with increasing
rate of Hubble parameter, negative deceleration parameter and
$\omega_{eff}<-1$. The trajectories of $r-s$ parameters identify
quintessence and phantom phases as $s>0$ when $r<1$. For
quintessence and phantom models, the scalar field continuously
increasing negatively and potential energy of the field is
dominating over kinetic energy. This analysis indicates that an
epoch of accelerated expansion is achieved for non-dust
distribution.

Shamir \cite{12} investigated exact solution of BI model without
using Noether symmetry approach in $f(R,T)$ gravity. For indirect
curvature-matter coupling, the exact solution is determined using a
relationship between expansion and shear scalars. The study of
corresponding cosmological parameters yield positive deceleration
parameter, $\omega_{eff}=1$, volume and average scale factor turn
out to be zero at $t=0$. Thus, the analysis of this exact solution
yields a decelerating epoch for $R+2f(T)$ model. For $f_1(R)+f_2(T)$
model, a power-law form of $f_1(R)$ is considered that gives
exponential and power-law solutions for different choices of
$f_2(T)$. For exponential solution, the average Hubble parameter
becomes zero leading to Einstein universe. Camci et al. \cite{k2}
formulated exact solutions of these anisotropic models via Noether
symmetry approach in non-minimally scalar coupled gravity. The scale
factors are found to be proportional to inverse of the scalar field
whose explicit form is not determined for any anisotropic model.
Consequently, the cosmological analysis of these exact solutions is
not established. In the present paper, we have found two exact
solutions, power-law and oscillatory solutions via Noether symmetry
approach that correspond to decelerating as well as current
accelerating universe for dust and non-dust distribution.

We conclude that the constructed $f(R,T)$ models admit direct as
well as indirect curvature-matter coupling. The existence of
symmetry generators and associated conserved quantities is assured
for both $f(R,T)$ models. It is worthwhile to mention here that we
have found maximum symmetry generators along with conserved
quantities for the second $f(R,T)$ model in the presence of boundary
term. This indicates that the model appreciating direct
curvature-matter coupling leads to more physical results relative to
the first model while the exact solutions analyze cosmic evolution.

\vspace{0.25cm}

{\bf Acknowledgment}

\vspace{0.25cm}

This work has been supported by the \emph{Pakistan Academy of
Sciences Project}.

\vspace{0.3cm}

\renewcommand{\theequation}{A\arabic{equation}}
\setcounter{equation}{0}
\section*{Appendix A}

For invariance condition (\ref{10}), the system of equations is
\begin{eqnarray}\label{6'}
&&\epsilon ab^2\eta,_{_t}=-B,_{_\phi},\\\label{a'}&&b\alpha
+2a\beta+2ab\eta,_{_\phi}-ab\tau,_{_t}=0,
\\\label{20'}&&2b\alpha,_{_\phi}f_{_RR}
+4a\beta,_{_\phi}f_{_RR}+ab\epsilon\eta,_{_R}=0,\\\label{21'}&&2b\alpha,_{_\phi}
f_{_RT}+2a\beta,_{_\phi}f_{_RT}+ab\epsilon\eta,_{_T}=0,\\\label{10'}&&4\beta,_{_\phi}
f_{R}+2b\gamma,_{_\phi}f_{RR}+2b\delta,_{_\phi}f_{RT}+ab\epsilon\eta,_{_a}=0,
\\\label{15'}&&4b\alpha,_{_\phi}f_{R}+4a\beta,_{_\phi}
f_{R}+4ab\gamma,_{_\phi}f_{RR}+4ab\delta,_{_\phi}f_{RT}+ab^2\epsilon\eta,_{_b}=0,
\\\label{1'}&&\tau,_{_a}f_R=0,\quad\tau,_{_b}f_R=0,\quad\tau,_{_R}f_{RR}=0,\quad
\tau,_{_T}f_{RT}=0,\quad\tau,_{_\phi}=0,\\\label{4'}&&2b^2\alpha,_{_t}f_{RR}
+4ab\beta,_{_t}f_{RR}=-B,_{_R},\\\label{5'}&&2b^2\alpha,_{_t}f_{RT}
+4ab\beta,_{_t}f_{RT}=-B,_{_T},\\\label{16'}&&b\alpha,_{_R}f_{RR}
+2ab\beta,_{_R}f_{RR}=0,\\\label{17'}&&b\alpha,_{_T}f_{RT}+2ab\beta,_{_T}f_{RT}=0,
\\\label{11'}&&2\beta,_{_a}f_R
+b\gamma,_{_a}f_{RR}+b\delta,_{_a}f_{RT}=0,
\\\label{2'}&&4b\beta,_{_t}f_{R}+2b^2\gamma,_{_t}f_{RR}
+2b^2\delta,_{_t}f_{RT}=-B,_{_a},
\\\label{3'}&&4b\alpha,_{_t}f_{R}+4a\beta,_{_t}f_{R}+4ab\gamma,_{_t}f_{RR}
+4ab\delta,_{_t}f_{RT}=-B,_{_b},\\\label{19'}&&b\alpha,_{_T}f_{RR}
+b\alpha,_{_R}f_{RT}+2a\beta,_{_T}f_{RR}+2a\beta,_{_R}f_{RT}=0,
\\\nonumber
&&\alpha f_R+a\gamma f_{RR}+a\delta
f_{RT}+2b\alpha,_{_b}f_R+2a\beta,_{_b}f_R+2ab\gamma,_{_b}f_{RR}
+2ab\delta,_{_b}f_{RT}\\\label{12'}&&-a\tau,_{_t}f_R=0,
\\\nonumber&&2\beta
f_{RR}+b\gamma f_{RRR}+b\delta
f_{RRT}+b\alpha,_{_a}f_{RR}+2a\beta,_{_a}f_{RR}+2\beta,_{_R}f_R
+b\gamma,_{_R}f_{RR}\\\label{8'}&&+b\delta,_{_R}f_{RT}-b\tau,_{_t}f_{RR}=0,
\\\nonumber&&2\beta
f_{RT}+b\gamma f_{RRT}+b\delta
f_{RTT}+b\alpha,_{_a}f_{RT}+2a\beta,_{_a}f_{RT}+2\beta,_{_T}f_R
+b\gamma,_{_T}f_{RR}\\\label{9'}&&+b\delta,_{_T}f_{RT}-b\tau,_{_t}f_{RT}=0,
\\\nonumber&&2\beta
f_R+2b\gamma f_{RR}+2b\delta
f_{RT}+2b\alpha,_{_a}f_R+4a\beta,_{_a}f_R+2b\beta,_{_b}f_R+2ab
\gamma,_{_a}f_{RR}\\\label{7'}&&+b^2\gamma,_{_b}f_{RR}+2ab\delta,_{_a}f_{RT}
+b^2\delta,_{_b}f_{RT}-2b\tau,_{_t}f_{R}=0,
\\\nonumber&&2b\alpha f_{RR}+2a\beta
f_{RR}+2ab\gamma f_{RRR}+2ab\delta
f_{RRT}+b^2\alpha,_{_b}f_{RR}+2b\alpha,_{_R}f_R+2ab\\\label{13'}&&
\times\beta,_{_b}f_{RR}+2a\beta,_{_R}f_R+2ab\gamma,_{_R}f_{RR}
+2ab\delta,_{_R}f_{RT}-2ab\tau,_{_t}f_{RR}=0,\\\nonumber&&2b\alpha
f_{RT}+2a\beta f_{RT}+2ab\gamma f_{RRT}+2ab\delta
f_{RTT}+b^2\alpha,_{_b}f_{RT}+2b\alpha,_{_T}f_R+2ab\\\label{14'}&&\times\beta,_{_b}f_{RT}
+2a\beta,_{_T}f_R+2ab\gamma,_{_T}f_{RR}+2ab\delta,_{_T}f_{RT}-2ab\tau,_{_t}f_{RT}=0,
\\\nonumber
&&b^2\alpha[f-Rf_R+f_T(3p_m-\rho_m-T)+p_m-V(\phi)+a\{f_T(3p_m,_{_a}-\rho_m,_{_a})
\\\nonumber&&+p_m,_{_a}\}+2\xi f_R]+\beta[2ab(f-Rf_R+f_T(3p_m-\rho_m-T)
+p_m-V(\phi))\\\nonumber&&+ab^2\{f_T(3p_m,_{_b}-\rho_m,_{_b})+p_m,_{_b}\}]
+\gamma[-ab^2Rf_{RR}+2a\xi
f_{RR}]+\delta[-ab^2Rf_{RT}\\\nonumber&&+2a\xi
f_{RT}]-ab^2V'(\phi)\eta
+\tau,_{_t}[ab^2(f-Rf_R+f_T(3p_m-\rho_m-T)+p_m\\\label{22'}&&-V(\phi))+2a\xi
f_R]=B,_{_t}.
\end{eqnarray}

\end{document}